\documentclass[fleqn,10pt,twocolumn]{wlscirep}
\usepackage{placeins}
\usepackage{textcomp}
\usepackage{hyperref}
\usepackage{float}
\usepackage{parskip}
\usepackage{siunitx}
\usepackage{hyperref}
\definecolor{mygreen}{rgb}{0.01, 0.5, 0.01}
\definecolor{myred}{rgb}{0.8, 0.4745098039215686, 0.6549019607843137}
\hypersetup{
    colorlinks=true,
    linkcolor=mygreen,
    citecolor=myred
}

\title{
Patchy landscapes promote stability of small groups
}

\author[a, *]{Gianni Jacucci}
\author[b]{Davide Breoni}
\author[c]{Sandrine Heijnen}
\author[d]{José Palomo}
\author[e]{Philip Jones}
\author[b]{Hartmut Löwen}
\author[c]{Giorgio Volpe}
\author[a]{Sylvain Gigan} 
\affil[a]{Laboratoire Kastler Brossel, Sorbonne Université, Ecole Normale Supérieure-Paris Sciences et Lettres (PSL) Research University, Centre National de la Recherche Scientifique (CNRS) UMR 8552, Collège de France, 24 rue Lhomond, 75005 Paris, France}
\affil[b]{Institut für Theoretische Physik II—Weiche Materie, Heinrich-Heine-Universität Düsseldorf, Universitätsstraße 1, D-40225 Düsseldorf, Germany}
\affil[c]{Department of Chemistry, University College London, 20 Gordon Street, London WC1H 0AJ, United Kingdom}
\affil[d]{Laboratoire de Physique de l’Ecole Normale Supérieure, ENS, Université PSL, CNRS, Sorbonne Université, Université Paris Cité, F-75005 Paris, France}
\affil[e]{Department of Physics $\&$ Astronomy, University College London, Gower Street, London WC1E 6BT, United Kingdom}
\affil[*]{To whom correspondence should be addressed. E-mail: giovanni.iacucci@lkb.ens.fr}

\begin{abstract}
Group formation and coordination are fundamental characteristics of living systems, essential for performing tasks and ensuring survival. Interactions between individuals play a key role in group formation, and the impact of resource distributions is a vibrant area of research. Using active particles in a tuneable optical environment as a model system, we demonstrate that heterogeneous energy source distributions result in smaller, more stable groups with reduced individual exchange between clusters compared to homogeneous conditions.
\textcolor{black}{Reduced group sizes can be beneficial to optimise resources in heterogeneous environments and to control information flow within populations. Devoid of biological complications, our system provides insights into the importance of patchy landscapes in ecological dynamics and holds implications for refining swarm intelligence algorithms and enhancing crowd control techniques.}
\end{abstract}

\begin{document}
\maketitle

\section*{Introduction}
Group living is widespread across different levels of biological organisation \cite{sumpterCollectiveAnimalBehavior2010} and confers many advantages to individuals, such as increased survival from predators \cite{doi:10.1073/pnas.1420068112, doi:10.1073/pnas.1905585116, klamserCollectivePredatorEvasion2021}, enhanced foraging efficiency \cite{brownFoodsharingSignalsSocially1991,jacksonCommunicationAnts2006, feinermanPhysicsCooperativeTransport2018}, or improved communication and decision making \cite{strandburg-peshkinSharedDecisionmakingDrives2015, doi:10.1126/science.1210280, brushConflictsInterestImprove2018}. 
The formation, maintenance, and disbanding of groups depend not only on the interactions among individuals but also on the characteristics of the surrounding environment, such as spatial or temporal heterogeneity \cite{gordonEcologyCollectiveBehavior2014}.
For instance, while large group sizes are advantageous for foraging in homogeneous environments, the presence of a patchy landscape of resources compels individuals to adopt smaller group formations as an adaptive strategy to ensure sufficient resources for all members \cite{gueronDynamicsGroupFormation1995, chapmanEcologicalConstraintsGroup1995, aureliFissionFusionDynamics2008, teichroebTestEcologicalconstraintsModel2009}.
However, the challenge of simultaneously monitoring multiple individuals in natural ecosystems poses limits to our understanding of the dynamics of such groups, particularly regarding their durability and time stability \cite{hugheyChallengesSolutionsStudying2018,tuiaPerspectivesMachineLearning2022}.

Recently, active colloids have emerged as a useful tool for studying collective behaviour in living systems \cite{romanczukActiveBrownianParticles2012,bechingerActiveParticlesComplex2016}.
These artificial systems, unlike natural ones \cite{cavagnaBirdFlocksCondensed2014,santiniTetraDENSITYDatabasePopulation2018,kingRewildingCollectiveBehaviour2018,northrupConceptualMethodologicalAdvances2022}, have the advantage of evolving on short timescales. Moreover, they are tuneable, and their small sizes make data acquisition more convenient. Previous studies have investigated the role of individual interactions in group formation, such as those due to attraction \cite{gregoireOnsetCollectiveCohesive2004,palacciLivingCrystalsLightActivated2013,ginotAggregationfragmentationIndividualDynamics2018}, repulsion \cite{buttinoniDynamicalClusteringPhase2013}, alignment \cite{vicsekNovelTypePhase1995, bricardEmergenceMacroscopicDirected2013}, reorientation \cite{vandrongelenCollectiveDynamicsSoft2015, zhangActivePhaseSeparation2021}, or vision \cite{durveActiveParticleCondensation2018, lavergneGroupFormationCohesion2019} and communication \cite{altemoseChemicallyControlledSpatiotemporal2017, ziepkeMultiscaleOrganizationCommunicating2022}.
The role of physical features in the environment has also been explored, such as the presence of obstacles \cite{PhysRevE.90.012701, morinDistortionDestructionColloidal2017, reichhardtCloggingDepinningBallistic2018, frangipaneInvariancePropertiesBacterial2019,kurzthalerGeometricCriterionOptimal2021,diasEnvironmentalMemoryBoosts2023} or of a disordered potential \cite{pinceDisordermediatedCrowdControl2016}.
However, the impact that a heterogeneous distribution of energy sources has on collective phenomena is not clear yet. Indeed, while work exists that studied the impact of linear gradients in the energy source \cite{lozanoPhototaxisSyntheticMicroswimmers2016, gomez-solanoTuningMotilityDirectionality2017, jahanshahiRealizationMotilitytrapActive2020}, natural landscapes of resources hardly follow linear gradients and the impact of patchiness on group dynamics is largely unknown.

Here, we use phototactic active colloids moving in a spatially complex distribution of resources, generated optically, to investigate the role of patchy landscapes in the dynamics of group formation and cohesion. 
Our results demonstrate that patchiness restricts the size of groups and increases their stability by decreasing the exchange of individuals compared to a homogeneous landscape. 
\section*{Results}

\begin{figure*}[t!]
    \centering
    \includegraphics[width=\linewidth, page=3]{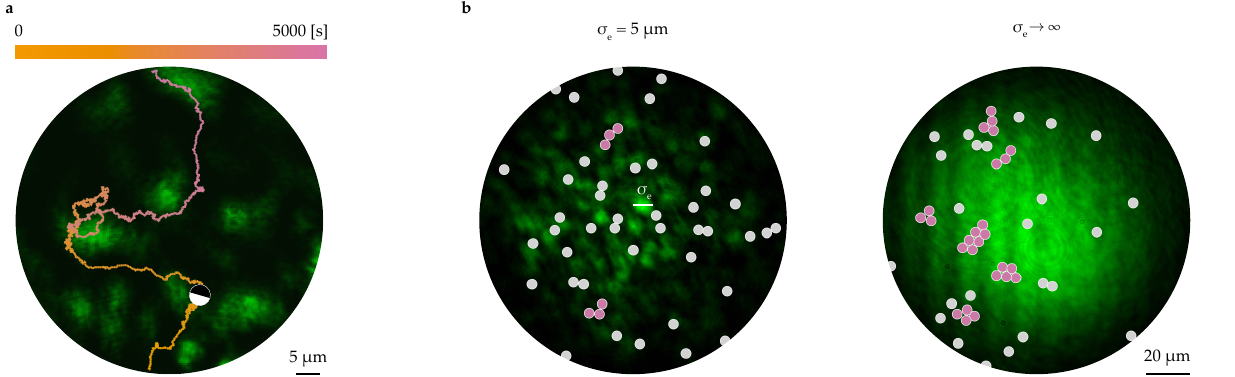}
    \caption{\textbf{Active particles in complex energy landscapes.} 
\textbf{a)} Trajectory of an active particle in a spatially complex distribution of energy, with high light intensities in green. The particle self-propels, avoiding the most intense illumination. The trajectory displays the particle's centre position over 45 minutes.
\textbf{b)} As the particles prefer to stay in the dark areas, the group (defined as comprising more than two individuals) size appears to be determined by the spatial characteristics of the environment—$\sigma_\text{e}$, i.e. the average size of the energy patches. The particles belonging to a group are coloured in pink while the isolated particles and the particles in pairs are in grey.}
    \label{fig1}
\end{figure*}

As phototactic active particles, we used monodispersed silica colloids (diameter $d = \SI{4.77 \pm 0.20}{\micro \meter}$) with a \SI{90}{\nano  \meter} carbon half-coating (Methods). These particles self-propel in a water-2,6-lutidine critical mixture when exposed to light. 
The propulsion arises from the critical mixture's local demixing, induced by the carbon heating upon light absorption.
\cite{volpeMicroswimmersPatternedEnvironments2011, buttinoniActiveBrownianMotion2012, buttinoniDynamicalClusteringPhase2013,lozanoPhototaxisSyntheticMicroswimmers2016,gomez-solanoTuningMotilityDirectionality2017,jahanshahiRealizationMotilitytrapActive2020,lozanoActiveBrownianParticles2022}.

We created an energy source landscape with controllable spatial heterogeneity by shining a laser on an optical diffuser, generating a diffraction pattern of randomly distributed light patches, \textcolor{black}{whose intensity follows a negative exponential probability distribution \cite{volpeBrownianMotionSpeckle2014, bianchiActiveDynamicsColloidal2016}.}
\begin{figure}[b!]
\vspace{-0.8cm}
    \centering
    \includegraphics[width=\linewidth, page=8]{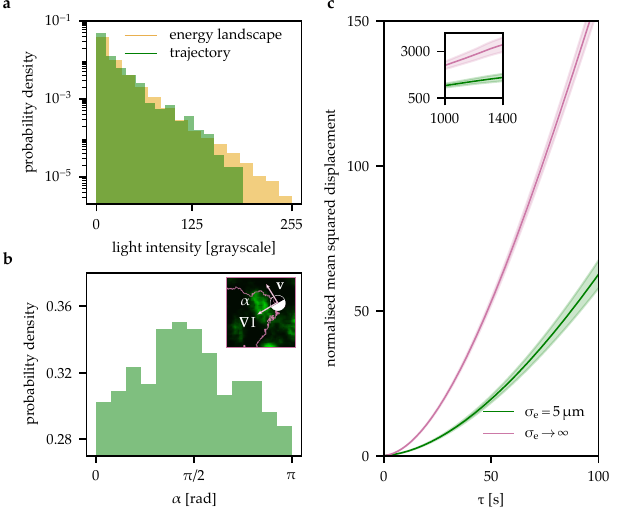}
    \caption{\textbf{Statistics of single particle trajectories on the energy landscape.} 
A complex energy landscape ($\sigma_\text{e}=\SI{5}{\micro \meter}$) affects the spatial occupancy and statistical features of particle trajectories. 
\textbf{a)} Histogram comparing the distribution of the light intensity in the locations visited by the particles with the intensity distribution of the energy landscape. 
\textbf{b)} Particles tend to move perpendicular to the gradient of the illumination ($\alpha=\pi /2$). $\alpha$, illustrated in the inset, is the angle between the velocity of the particle (\textbf{v}) and the intensity gradient ($\mathbf {\nabla}I$).
\textbf{c)} A lower mean squared displacement reflects avoidance of high-energy areas compared to a more homogeneous environment (Gaussian illumination, $\sigma_\text{e} \to \infty$). \textcolor{black}{The inset shows the long-time diffusive behaviour.}
Shaded areas indicate standard deviations on the mean squared displacement. Data was obtained from averaging at least 20 trajectories, each lasting 45 minutes.}
    \label{fig2}
\end{figure}
The energy density of the landscape and the typical size of its patches ($\sigma_\text{e}$) can be precisely controlled by optical means—\autoref{figS1} and Methods for details. 
\textcolor{black}{$\sigma_\text{e}$ is defined as the full width at half maximum of the landscape's autocorrelation}. 
\autoref{fig1}\textcolor{mygreen}{a} shows a typical example of a particle trajectory in a complex illumination. 
What can be qualitatively observed is that an active particle tends to navigate in the energy landscape by spending most of its time in the low-illumination areas—in other words, it performs negative phototaxis. 
This property of the individual trajectories has a direct consequence on group formation, with a group being defined as a collection of more than two particles in contact.
As shown in \autoref{fig1}\textcolor{mygreen}{b}, in a heterogeneous energy landscape ($\sigma_\text{e} \simeq d = \SI{5}{\micro \meter}$), active particles tend to form smaller groups compared to the case of a homogeneous illumination ($\sigma_\text{e} \to\infty$).
\textcolor{black}{Under both illumination conditions, the average local energy density ($I =  \SI{0.1}{\micro \watt \micro \meter^{-2}}$) was kept the same by matching the incoming power and envelope size \autoref{figS5}}.

\begin{figure*}[t!]
    \centering
    \includegraphics[width=\linewidth, page=2]{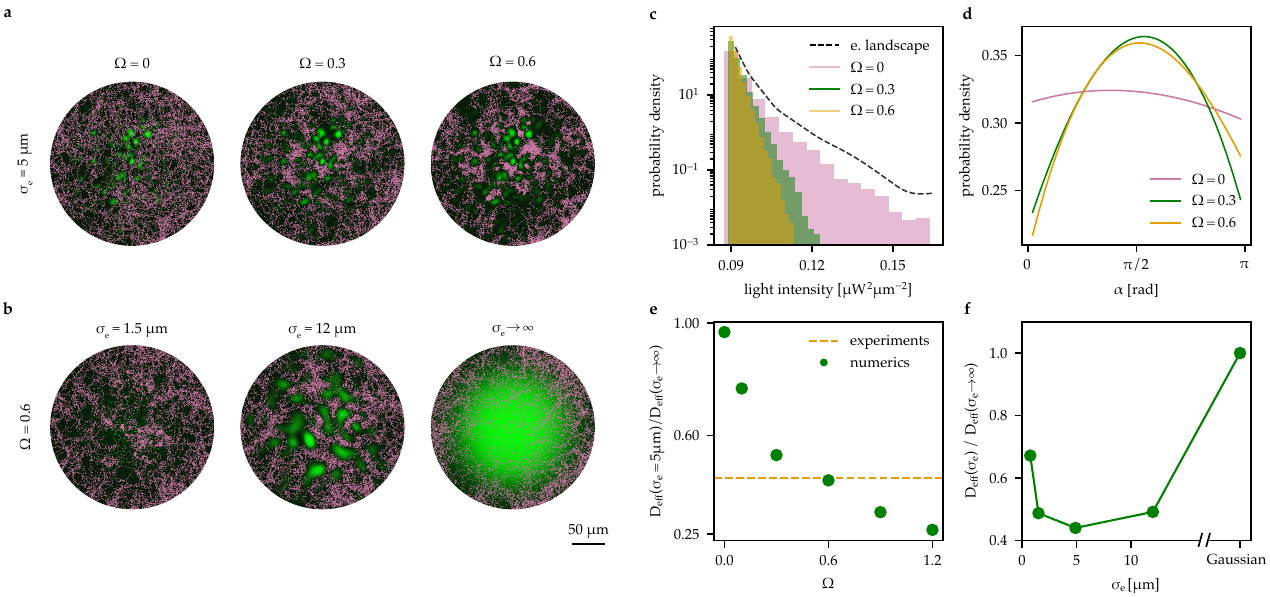}
    \caption{\textbf{Modeling the dynamics of active particles in complex energy landscapes.} 
    A torque and spatial heterogeneities in the energy landscape determine the dynamics of the active particles.
    Simulated trajectories of individual active particles with \textbf{a)} different torque values ($\Omega$) for a fixed $\sigma_\text{e}=\SI{5}{\micro \meter}$, and \textbf{b)} environment heterogeneity ($\sigma_\text{e}$), for a fixed $\Omega=6$. 
    \textbf{c)} In a complex environment ($\sigma_\text{e}=\SI{5}{\micro \meter}$), increasing the torque causes the particles to avoid high-energy areas. \textbf{d)} In the presence of a torque, particles move more perpendicular to the gradient ($\alpha=\pi/2$). 
    \textbf{e)} \textcolor{black}{The difference in the effective diffusion coefficients between complex and Gaussian illuminations observed in the experiments is most accurately described by $\Omega=0.6$.}
    \textbf{f)} Role of the heterogeneity of the energy landscape ($\sigma_\text{e}$) on the effective diffusion coefficient ($D_\text{eff}(\sigma_e)$) for a fixed value of the torque ($\Omega=0.6$). The effective diffusion coefficient, normalised to its value for a Gaussian illumination, is minimised when the spatial variation of the landscape is comparable to the size of the particles. All simulations averaged 150 particles over 50 minutes each. 
    \textcolor{black}{The standard deviation of the data is represented in e) as shaded areas for experiments and vertical bars for numerics, comparable with the size of the scatter plot points. In f), the errors match the scatter point size.}
}
\label{fig3sim}
\end{figure*}

In order to explain the phenomena illustrated in \autoref{fig1}, we first need to better understand the statistics of the dynamics of the individual particles on the underlying energy landscape.  
\textcolor{black}{
\autoref{fig2}\textcolor{mygreen}{a} shows quantitatively that particles perform negative phototaxis. By comparing the light intensity at locations visited by the particles along their trajectories with the overall energy landscape, we can observe that particles tend to avoid the most intense light regions.
Delving deeper, \autoref{fig2}\textcolor{mygreen}{b} shows that as active particles navigate the darker, low-energy channels of the light landscape, they tend to align perpendicular to the light intensity gradient. This is due to their continuous interactions with the two-dimensional gradient, channelling their trajectories between high-energy patches and causing them to move perpendicular to the gradient—i.e. along low-intensity areas.} 
This is evidenced by the peak at the angle $\alpha=\pi/2$ in \autoref{fig2}\textcolor{mygreen}{b}. Here, $\alpha$ represents the angle between the particle's instantaneous velocity and the local intensity gradient, as illustrated in the inset of \autoref{fig2}\textcolor{mygreen}{b}.
This result extends beyond observations in one-dimensional light gradients, where previous findings associated $\alpha=\pi$ ($\alpha=0$) with negative (positive) phototaxis \cite{jahanshahiRealizationMotilitytrapActive2020}. 

A complex energy landscape determines not only the preferential orientation of the active colloids but also their speed.
Both in homogeneous energy landscapes \cite{lozanoPhototaxisSyntheticMicroswimmers2016, gomez-solanoTuningMotilityDirectionality2017} and in one-dimensional light gradients \cite{jahanshahiRealizationMotilitytrapActive2020} a linear increase of the velocity with the light intensity was observed after a given activation energy. 
This is also reproduced in two-dimensional gradients when the illumination has a Gaussian profile ($\sigma_\text{e} \to \infty$ in \autoref{figS4}\textcolor{mygreen}{a}). In this case the local value of the gradient is relatively low, making it comparable to the previously reported linear cases also in terms of orientation with respect to the gradient ($\alpha=\pi$ in \autoref{fig3}\textcolor{mygreen}{a}) \cite{lozanoPhototaxisSyntheticMicroswimmers2016, gomez-solanoTuningMotilityDirectionality2017, jahanshahiRealizationMotilitytrapActive2020}. 
However, when the heterogeneity of the landscape is comparable to the size of the particles ($\sigma_\text{e}=\SI{5}{\micro \meter}$ in \autoref{figS4}\textcolor{mygreen}{b}) the velocity does not show a monotonic increase with the light intensity. This results from local values of the light gradient which are much stronger than in the previous cases, giving rise to a complex dependency of the speed of the particle on the energy landscape. 
Active particles do indeed experience a torque proportional to the light gradient, which, in turn, is proportional to the local light intensity. This torque causes them to change direction and reduce speed near high-energy patches \cite{lozanoPhototaxisSyntheticMicroswimmers2016, gomez-solanoTuningMotilityDirectionality2017, jahanshahiRealizationMotilitytrapActive2020}. 
The aligning torque is a consequence of breaking the axial symmetry of the velocity field around the particle \cite{bickelPolarizationActiveJanus2014a}. 
The torque-induced reorientation and avoidance of high-energy zones reduce the mean squared displacement of active particles in a complex landscape \textcolor{black}{ when compared to a Gaussian illumination,} as seen in \autoref{fig2}\textcolor{mygreen}{c}. The mean squared displacement, despite the decrease, exhibits a superdiffusive transition followed by a long-time return to typical diffusion, as for typical active colloids in both scenarios \cite{bechingerActiveParticlesComplex2016}.

To disentangle the role of the torque (the dimensionless parameter $\Omega$, \autoref{fig3sim}\textcolor{mygreen}{a}) exerted by the field gradient and the heterogeneity of the energy landscape ($\sigma_\text{e}$, \autoref{fig3sim}\textcolor{mygreen}{b}) on the microscale dynamics, we employed a particle-based model that includes an aligning torque proportional to the gradient of the optical field (Methods) \cite{lozanoPhototaxisSyntheticMicroswimmers2016, gomez-solanoTuningMotilityDirectionality2017, jahanshahiRealizationMotilitytrapActive2020}. \autoref{fig3sim}\textcolor{mygreen}{c} confirms how the presence of a torque prevents particles from accessing the high energy regions, with a cutoff in energy that decreases as the torque increases.  The emergence of this behaviour coincides also with a preferential orientation of the particles' velocity perpendicular to the gradient of the illumination, whereas at $\Omega=0$ the velocity shows a relatively weak dependence on the gradient (\autoref{fig3sim}\textcolor{mygreen}{d}). 
Our model further validates that under a two-dimensional Gaussian illumination, the preferred orientation relative to the gradient ($\alpha=\pi$ in \autoref{fig3}\textcolor{mygreen}{b}) is governed by the effect of the torque.
\textcolor{black}{
Additionally, as depicted in \autoref{fig3sim}\textcolor{mygreen}{e}, the discrepancy in the long-term behavior of the mean squared displacement between $\sigma_\text{e}=\SI{5}{\micro \meter}$ and $\sigma_\text{e} \to \infty$ in \autoref{fig2} can be attributed to the torque. In a complex illumination, an increase in torque restricts access to regions of higher energy, thereby reducing the mean squared displacement.
}
Importantly, altering the heterogeneity of the environment (and thus the gradient value) while keeping the torque constant also results in a lower mean squared displacement, as shown in \autoref{fig3sim}\textcolor{mygreen}{f} (represented as effective diffusion coefficient). This finding highlights the significant influence of environmental heterogeneity on particle dynamics— i.e. when $\sigma_\text{e}$ is on a similar length scale to the particle size, it significantly affects its long-time diffusion coefficient. 
Conversely, in the asymptotic cases where energy variations occur on scales much smaller or larger than the particle, differences on long-time diffusivity are less marked.

To study the impact of the differences in individual trajectories on the collective behaviour of active systems we performed experiments with a higher concentration of particles ($\approx 4 \%$, compared to $\approx 0.1 \%$ in \autoref{fig2}).
When the concentration of active particles increases, they are more likely to collide with each other, slowing down and forming groups. A group lasts until the propulsion direction of one of the individuals points outside the cluster and its speed is such that it can escape the group by winning over the short-range attraction among particles \cite{buttinoniDynamicalClusteringPhase2013,ginotAggregationfragmentationIndividualDynamics2018}. 

\begin{figure}[t!]
    \centering
    \includegraphics[width= \linewidth, page=9]{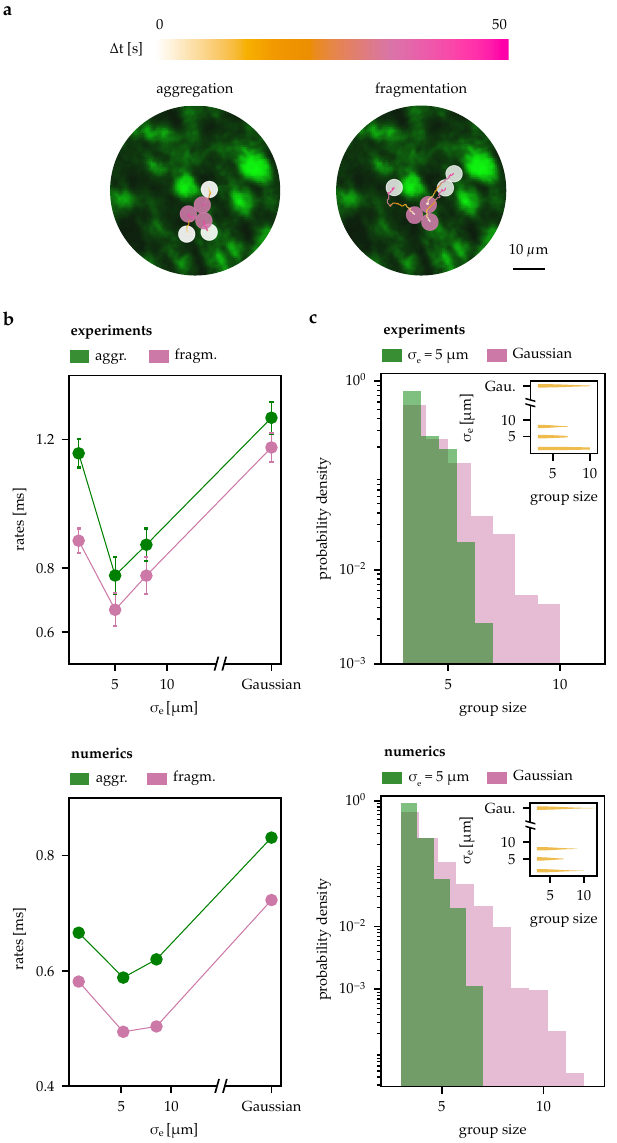}
    \caption{\textbf{Group formation analysis.} 
    Comparison of group formation in energy landscapes with different spatial characteristics, i.e. different sizes of the energy patches ($\sigma_\text{e}$).
    \textcolor{black}{
    \textbf{a)} Experimental trajectories showing an example of group aggregation and fragmentation in a complex landscape with $\sigma_\text{e}=\SI{5}{\micro \meter}$.
    When the patchiness of the landscape is comparable to the dimension of the particles, it reduces both the rate of individual exchange between them and the group size, \textbf{b)} and \textbf{c)}, respectively.
    In both panels of \textbf{c)}, the insets show the distribution of sizes for different spatial heterogeneities. 
    In b), the error bars represent the standard deviation of the data.}
    }
    \label{fig4}
\end{figure}

A heterogeneous energy landscape plays a key role in the formation of groups. 
As illustrated in \autoref{fig4}\textcolor{mygreen}{a}, particles tend to aggregate, and eventually separate, in the dark regions of the optical landscape where their motility is lower. 
In complex energy landscapes, groups exhibit greater durability, with the average rate of aggregation and fragmentation events minimised when the spatial heterogeneity matches the individual particle scale (\autoref{fig4}\textcolor{mygreen}{b}).
The torque does indeed play a dual role in maintaining group stability. Firstly, it makes encounters between individuals more difficult by causing particles to linger in low energy regions for longer periods of time, which reduces the rate of aggregation. Secondly, once groups have formed, the torque orients particles toward areas of low energy, causing them to face inward the cluster and reducing the likelihood of fragmentation.
The importance of the torque is further confirmed by \autoref{figS6}\textcolor{mygreen}{a}, where decreasing it leads to i) an overall decrease of the rates and ii) a smaller difference in rates between homogeneous and complex energy landscapes.

The patchiness of the landscape also has a significant impact on group size, as demonstrated in \autoref{fig4}\textcolor{mygreen}{c}.
As individuals display avoidance behaviour towards high energy regions, groups tend to grow until they fill the space confined between adjacent high-energy patches. The data presented in \autoref{fig4}\textcolor{mygreen}{c} support this qualitative relationship, revealing that complex energy landscapes with heterogeneity comparable to the particle size lead to a reduction in the group dimension. The interplay between the dimension of energy patches and the group size is strongly influenced by the torque exerted by the gradient of resources. 
Decreasing the torque provides colloids with more freedom to move, enabling larger groups to form, as depicted in \autoref{figS6}\textcolor{mygreen}{a}. 

Notably, the differences in group dynamics depicted in \autoref{fig4} cannot be solely attributed to a lower \textcolor{black}{average motility of active particles in complex landscapes. 
Even when the mean squared displacement in a homogeneous energy landscape is reduced to match that in patchy ones (as shown in \autoref{figS6}\textcolor{mygreen}{b}), disparities persist in both group dimension and stability.} This further demonstrates that the spatial complexity of the energy landscape plays a critical role in shaping group dynamics.

\section*{Discussion}
In summary, our study investigated how heterogeneous energy landscapes affect group formation using active particles in optical fields as a model system. We found that by manipulating the spatial complexity of the landscape, we can control the size of groups and the exchange of individuals between them. 
\textcolor{black}{Our findings offer important biological perspectives. While smaller groups optimise resource allocation in patchy environments, their increased stability and reduced individual exchange might compromise genetic diversity. Yet, such limited interchange could also curtail the spread of diseases and parasites among groups and the transmission of antibiotic resistance in bacteria \cite{bennett2008plasmid}. 
As our experimental approach provides a versatile and precise method to modulate the spatial dynamics of the energy landscape using light—with a natural extension to explore its temporal properties—it could be instrumental in studying search strategies in the presence of limited resources. Its applicability could span various contexts, ranging from animals engaged in foraging or migration to guiding robots through complex search-and-rescue missions.}

\section*{Methods}
\subsection*{Materials}
Glass capillaries were purchased from CM Scientific (5005-050 and 5001-050 for single particle and cluster experiments, respectively). The lutidine was purchased and used as received: 2,6-lutidine ($\geq$ 99\%, Sigma-Aldrich). Deionised (DI) water ($\geq$ \SI{18}{\mega \Omega \cdot\centi\meter}) was collected from a Milli-Q purification system. Aqueous colloidal dispersions (5\% w/v) of silica (SiO$_{\mathrm{2}}$) colloids, used for sample preparation, were purchased from Microparticles GmbH. Additionally, carbon rods measuring \SI{300}{\milli\meter} in length and \SI{6.15}{\milli\meter} in diameter were procured from Agar Scientific and subsequently cut to a length of \SI{50}{\milli\meter} and tapered for the coating of Janus particles through sputtering.

\subsection*{\label{sec:ParticlesPrep}Particles fabrication}
Janus particles were synthesised from SiO$_{\mathrm{2}}$ colloids with a diameter of $d = \SI{4.77 \pm 0.20}{\micro \meter}$. Initially, a monolayer of colloids was deposited on a clean glass slide by evaporating a \SI{40}{\micro \liter} droplet containing a 2.5\% w/v dispersion of the colloids in DI water. Subsequently, an automatic carbon coater (AGB7367A, Agar Scientific) was employed to coat the particles with a \SI{90}{\nano \meter} thick layer of carbon. The thickness of the carbon layer was confirmed through atomic force microscopy (AFM) measurements. After coating, sonication was performed to dislodge the half-coated particles in DI water from the glass slide, facilitating their use for sample preparation.
To reduce the interaction between particles, and the sticking of the particles to the substrate, the Janus colloids were functionalised with bovine serum albumin (BSA, Sigma Aldrich). This was done by replacing the water solvent of the colloidal suspension with a 1$\%$w/v water solution of BSA. Prior to utilisation, the colloidal dispersions underwent centrifugation at 1000 relative centrifugal force (RCF) for 3 minutes, resulting in the formation of a pellet. The supernatant was subsequently discarded and replaced with a solution containing 28.6$\%$ w/v of water and 2,6-lutidine. This purification procedure was repeated three times to ensure the removal of any remaining BSA water solution from the original dispersion.

\subsection*{\label{sec:ExperimentsPrep}Sample preparation}
In single-particle experiments (\autoref{fig1} and \autoref{fig2}), a suspension of Janus particles was confined in a rectangular capillary with a width of \SI{700}{\micro \meter}, a length of \SI{50}{\milli \meter}, and a thickness of \SI{50}{\micro \meter}. The capillary was then sealed by applying a UV-curable glue to avoid evaporation and drifts.
In the multiple-particles systems (\autoref{fig4}), we reduced the sample's thickness by using a capillary with a width of \SI{100}{\micro \meter}, a length of \SI{50}{\milli \meter}, and a thickness of \SI{10}{\micro \meter} to restrict the motion of the colloids to two dimensions.

\subsection*{\label{sec:ExperimentsSetup}Experimental setup}
\autoref{figS1} shows a schematic of the experimental setup. A laser (Oxxius \SI{532}{\nano \meter}, \SI{300}{\milli \watt} of maximum output power) and a diffuser (\SI{1}{\degree}, Newport 10DKIT-C1) are used to illuminate the Janus particles with a random optical field, also known as speckle. 
The laser is directed to the sample via two mirrors (M1, M2) and a 4f-lens configuration (L1=L2, Thorlabs, LA1461-A-ML) illuminating the back aperture of a 10x objective (Nikon, N10X-PF, NA=0.3).
By changing the position of the diffuser and that of the sample, it is possible to control both the dimension of the speckle grains and the size of the illumination envelope. To create a Gaussian illumination, we removed the diffuser and one lens (L1) to image the beam on the sample. The illumination envelope was fixed to be close to the acquisition area of the camera ($\approx$ \SI{200 \times 200}{\micro \meter \squared} and $\approx$ \SI{100 \times 100}{\micro \meter \squared} in the single and multiple particles experiments, respectively) and it was kept the same for different patchiness of the landscape (\autoref{figS5}). 
The intensity of the laser—controlled by a half-waveplate (FOCtek, WPF212H) and a polarising beam splitter (FOCtek, BSC1204) was then adjusted to have a comparable energy density in the different optical landscapes, i.e. different values of $\sigma_\text{e}$. The active particles were imaged with a 20x objective (Nikon, N20X-PF, NA=0.5) and a tube lens (L3, Thorlabs, LA1805-A-ML) on a camera (Basler, acA5472-17um). A white LED (Thorlabs, MWWHF2), for simplicity not depicted in \autoref{figS1}, coupled to a fibre (Thorlabs, M28L01) is used for imaging.
The particles' dynamics were analysed by reconstructing their trajectories from videos (typically of the duration of 50 minutes at 2 frame per second (fps)) using a homemade code based on the python package \textit{Trackpy} \cite{crockerMethodsDigitalVideo1996,allan_daniel_b_2023_7670439}.
The position of the sample was controlled in three dimensions using an \textit{xyz}-stage (Thorlabs, RB13M/M). The \textit{z}-coordinate was adjusted with a stepper motor (Thorlabs, ZFS25B).
For all experiments, the suspension of Janus particles was kept close to the critical temperature of the water-lutidine mixture ($T_{\text{exp}}\simeq \SI{31}{\celsius} $, $\Delta T \simeq \SI{3}{\celsius}$) by a heater driven via a temperature controller (Thorlabs, HT19R and TC200, respectively).

\subsection*{\label{sec:Numerics}Numerical model}
We numerically implemented an Euler-Maruyama time integrator for the evolution of the positions $\textbf{r}_i$ and orientations $\phi_i$ of each particle $i$ in two dimensions, defined by the following equations, based on \cite{jahanshahiRealizationMotilitytrapActive2020}:
\begin{align}
    \dot{\textbf{r}}_i(t) &= v(\textbf{r}_i)\textbf{n}_i-\frac{1}{\gamma '}\sum_{j\neq i}\nabla V\left(\left|\textbf{r}_i-\textbf{r}_j\right|\right)+\sqrt{2D}\boldsymbol{\xi}_i(t), \label{eq:position}\\
   \dot{\phi}_i(t) &= \omega(\phi_i,\textbf{r}_i)+\sqrt{2D_r}\xi_i^\phi (t), \label{eq:angle}
\end{align}

where $\textbf{n}_i=(\cos(\phi),\sin(\phi))$, $\boldsymbol{\xi}_i$ and $\xi_i^\phi$ represent Gaussian white noise. The translational and rotational diffusion coefficients $D$ and $D_\text{r}$ are defined as
\begin{equation}
    D=\frac{k_BT}{\gamma'},\quad D_\text{r}=\frac{k_BT}{\beta'},
\end{equation}
\textcolor{black}{where $\gamma'=\frac{16}{3}\pi\eta d$ and $\beta'=\frac{8}{7}\pi\eta d^3$ are respectively the translational and rotational friction coefficient, corrected by taking into account a distance from the substrate $s=d/2$ \cite{diasEnvironmentalMemoryBoosts2023}, and $\eta= \SI{2.1 \times 10^{-3}}{\pascal \second}$ is the viscosity of water-lutidine. The torque $\omega$, which is defined as:}
\begin{equation}
    \omega(\phi_i,\textbf{r}_i)=\frac{\textcolor{black}{\Omega d}}{D}v(\textbf{r}_i)\left[\nabla v(\textbf{r}_i) \times \textbf{n}_i\right]\cdot \hat{e}_z,
\end{equation}
where the adimensional prefactor $\Omega$ determines whether the torque steers the particle towards the light ($\Omega<0$) or away from it ($\Omega>0$). 
\textcolor{black}{The value of $\Omega$ that best fits the experimental change in the long-time mean squared displacement is $\Omega=0.6$ (\autoref{fig3sim}\textcolor{mygreen}{e}).}
$V$ is a 6-12 Lennard-Jones potential with cutoff distance at five colloid diameters $d$ and depth $\epsilon\simeq 8k_\text{B}T$. \textcolor{black}{This value was chosen by fitting the experimental cluster aggregation and fragmentation rates for a Gaussian illumination.}
The intensity speckle field was generated by scaling the Fourier transform of a white noise map to match the specific characteristics of the optical fields observed in experiments: average intensity per unit of area and grain size. This intensity field was then translated into a motility field using the results from \cite{jahanshahiRealizationMotilitytrapActive2020}. 
\textcolor{black}{The gradient of the motility field acting on the particle} $\nabla v(\Vec{r}_i)$ was averaged over the  square area where the particle is inscribed.

\subsection*{\label{sec:NumericsCluster}Cluster analysis}
We defined the clusters by measuring all distances among particles and then creating bonds between all the particles with distances smaller than the cutoff radius 1.25d. 
\textcolor{black}{Similarly to Ref. \cite{ginotAggregationfragmentationIndividualDynamics2018,diasEnvironmentalMemoryBoosts2023}, clusters made of 2 particles (i.e. dimers) were disregarded in the analysis as they are unstable in time}. We determined whether a cluster survives from one timestep to the next by confronting its particles with those of all the clusters in the next timestep. We consider any of these clusters to be the same as a previous one if it satisfies two properties: i) that more than half of its particles are shared with the old cluster (i.e. the cluster has not decreased more than half in size); ii) that at least half of the particles of the new cluster are also present in the old one, i.e. the cluster has not grown more than double in size.
We calculated mainly two properties regarding clusters for each experiment: the cluster size and the monomer acquisition/loss rate. The first is simply the average number of monomers for each cluster, while the second is defined as the number of monomer acquisition/loss events (i.e. the number of times clusters receive/lose a single monomer during the experiment) normalised by the total experiment time and the average number of particles.

\section*{Data availability}
All data needed to evaluate the conclusions in the paper are present in the paper and/or the Supplementary Materials.

\section*{Acknowledgements}
The authors thank Dr. Raphaël Jeanneret and Dr. Jean Francois Allemand for suggesting the use of bovine serum albumin to reduce particles attraction and Alessia Gentili for insightful discussions on cluster analysis.
This project has received funding from the European Union’s Horizon 2020 research and innovation programme under the Marie Skłodowska-Curie grant agreement No 945298-ParisRegionFP, GJ is a fellow of the Paris Region Fellowship Programme supported by the Paris Region. 
The work of HL was supported by the German Research Foundation (project LO 418/29-1).
DB and SH acknowledge support by the EU MSCA-ITN ActiveMatter (Proposal No. 812780).

\section*{Author Contributions}
Author contributions are defined based on the CRediT (Contributor Roles Taxonomy) and listed alphabetically. Conceptualization: SG, GJ, HL, GV. Data curation: DB, GJ. Formal analysis: DB, GJ. Resources: SG, SH, PJ, HL, JP, GV. Writing: GJ wrote the manuscript with the help of all authors. 
Funding acquisition: SG, GJ, HL, GV. 

\section*{Conflict of interest}
The authors declare that they have no competing financial interests.

\bibliography{ms}
\newpage
\textcolor{white}{newPage}
\newpage
\setcounter{figure}{0}
\renewcommand{\thefigure}{S\arabic{figure}}

\begin{figure*}[t]
    \centering
    \includegraphics[width= 0.5\linewidth, page=1]{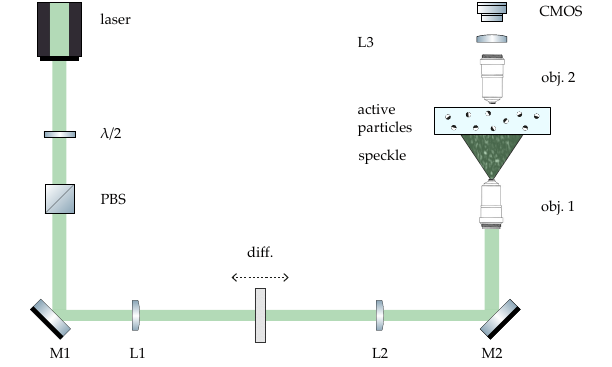}
    \caption{\textbf{Experimental setup.} 
    A laser and a diffuser are used to generate a spatially complex energy landscape for active particles.
    The laser is directed to the sample via two mirrors (M1, M2) and a 4f-lens configuration (L1=L2) illuminating the back aperture of an objective (obj. 1). Active particles are imaged via a second objective (obj. 2), a tube lens (L3) and a camera (CMOS). The trajectories are then reconstructed using a custom software.
    The speckle grain size and envelope can be adjusted by changing the longitudinal position of the diffuser with respect to the sample. The laser power is controlled by a half-waveplate ($\lambda$/2) and a polarising beamsplitter (PBS).
   }
    \label{figS1}
\end{figure*}

\begin{figure*}[t]
    \centering
    \includegraphics[width= 0.5\linewidth, page=10]{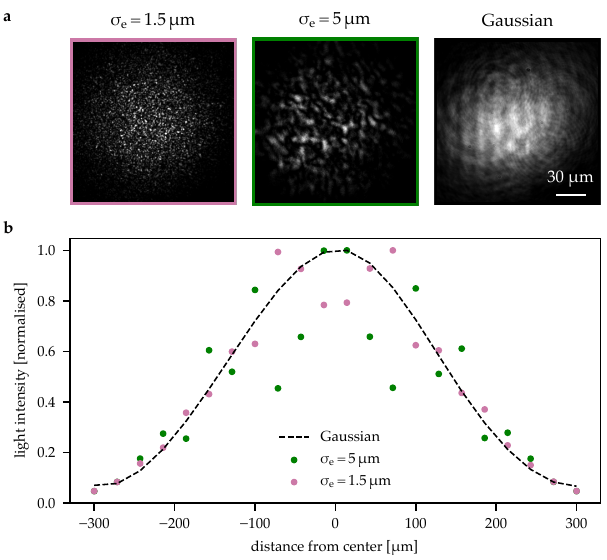}
    \caption{\textbf{Illumination profile.} 
    \textbf{a)} Three representative illumination patterns for different speckle grain sizes ($\sigma_e$). 
    \textbf{b)} Radially averaged intensity profile as a function of distance, based on images in a). 
    The illumination profiles share a common envelope, indicating that the active particles experience the same energy density across different energy landscapes.}
    \label{figS5}
\end{figure*}

\begin{figure*}[t]
    \centering
    \includegraphics[width= 0.5\linewidth, page=7]{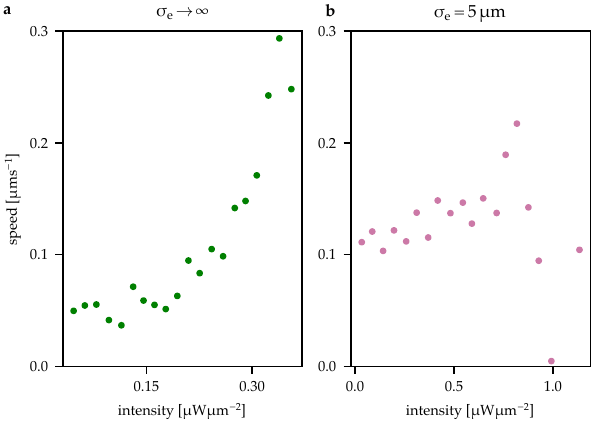}
    \caption{\textbf{Velocity as a function of light intensity.} 
    \textbf{a)} In a Gaussian illumination ($\sigma_\text{e} \to \infty$), where the gradient is small, there is a monotonic increase in velocity with light intensity. 
    \textbf{b)} In contrast, in a complex case ($\sigma_\text{e}=\SI{5}{\micro \meter}$)—where the local gradient is much larger and linearly dependent on the light intensity—the speed shows \textcolor{black}{a weak dependence on the intensity. This arises as a result of the torque exerted by the gradient, causing particles to change direction and therefore reducing their speed near high-intensity regions.}
    }    
    \label{figS4}
\end{figure*}

\begin{figure*}[t]
    \centering
    \includegraphics[width=.5\linewidth, page=4]{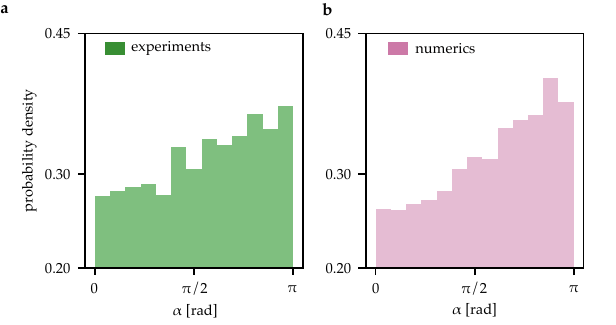}
    \caption{\textbf{Orientation of an active particle in a Gaussian energy landscape.} 
    When moving in a Gaussian energy landscape ($\sigma_\text{e} \to \infty$)—i.e. in the presence of a light gradient on a scale much larger than their size—active particles tend to align preferentially antiparallel to the global gradient of the illumination ($\alpha=\pi$).
    \textbf{a)} experiments, \textbf{b)} numerics.
    }
    \label{fig3}
\end{figure*}

\begin{figure*}[t]
    \centering
    \includegraphics[width= 0.5\linewidth, page=5]{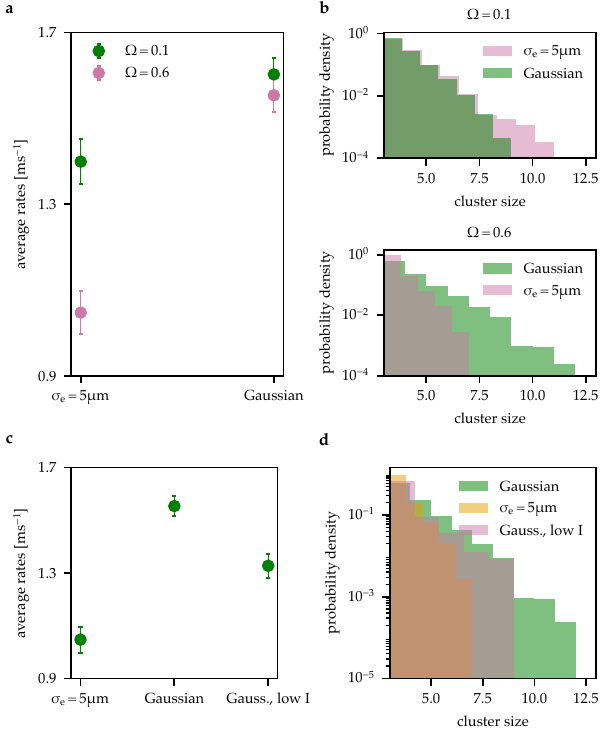}
    \caption{\textbf{Effect of torque and lower motility on group formation.} 
    Numerical simulations show that a smaller torque ($\Omega=0.1$ vs $\Omega=0.6$) leads to a less marked difference in rates \textbf{a)} and group size \textbf{b)} between complex ($\sigma_\text{e}=\SI{5}{\micro \meter}$) and Gaussian motility fields (Gaussian).
    In order to investigate whether the difference in group dynamics was solely influenced by the lower average motility of active particles in complex illumination, simulations were conducted in a homogeneous motility field with an identical average motility to that of the inhomogeneous scenario—achieved by reducing the light intensity (Gaussian, low I). However, reducing the average motility does not bridge the gap, in terms of both rates and sizes, \textbf{c)} and \textbf{d)}, respectively, between homogeneous and complex illuminations.}
    \label{figS6}
\end{figure*}
\end{document}